\documentclass[twocolumn]{aastex631mod}
\usepackage{amsmath,amssymb}
\usepackage{float}
\usepackage[nodayofweek]{datetime}
\newdateformat{monthyearday}{%
  \THEYEAR\ \monthname[\THEMONTH] \twodigit{\THEDAY}}

\makeatletter
\patchcmd{\NAT@citex}
  {\@citea\NAT@hyper@{%
      \NAT@nmfmt{\NAT@nm}%
      \hyper@natlinkbreak{\NAT@aysep\NAT@spacechar}{\@citeb\@extra@b@citeb}%
      \NAT@date}}
  {\@citea\NAT@nmfmt{\NAT@nm}%
    \NAT@aysep\NAT@spacechar\NAT@hyper@{\NAT@date}}{}{}

\patchcmd{\NAT@citex}
  {\@citea\NAT@hyper@{%
      \NAT@nmfmt{\NAT@nm}%
      \hyper@natlinkbreak{\NAT@spacechar\NAT@@open\if*#1*\else#1\NAT@spacechar\fi}%
        {\@citeb\@extra@b@citeb}%
      \NAT@date}}
  {\@citea\NAT@nmfmt{\NAT@nm}%
    \NAT@spacechar\NAT@@open\if*#1*\else#1\NAT@spacechar\fi\NAT@hyper@{\NAT@date}}
  {}{}
\makeatother

\newcommand{\med}[1]{\ensuremath{\left\langle\right.\!\!{#1}\!\!\left.\right\rangle}}
\newcommand{\age}{\ensuremath{\mathrm{age}}}
\newcommand{\tf}{\ensuremath{t_\mathrm{form}}}
\newcommand{\met}{\hbox{\ensuremath{\rm[Z/H]}}}
\newcommand{\afe}{\hbox{\ensuremath{\rm[\alpha/Fe]}}}
\newcommand{\kms}{\ensuremath{\rm km~s^{-1}}}
\newcommand{\kmsMpc}{\ensuremath{\rm km~s^{-1}~Mpc^{-1}}}
\newcommand{\Om}{\ensuremath{\Omega_{m,0}}}
\newcommand{\OL}{\ensuremath{\Omega_{\Lambda,0}}}
\defcitealias{Borghi2022a}{Paper~I}
\defcitealias{Thomas2011}{TMJ11}
\defcitealias{Vazdekis2015}{V15}

\accepted{December 3, 2021}

\shorttitle{A new measurement of $H(z)$ at $z\sim0.7$}
\shortauthors{Borghi et al. 2022b}
\graphicspath{{./}}

\begin{document}

\title{Toward a Better Understanding of Cosmic Chronometers:\\ A new measurement of \ensuremath{\boldsymbol{H(z)}} at \ensuremath{\boldsymbol{z\sim0.7}}}

\author[0000-0002-2889-8997]{Nicola Borghi}
\affiliation{Dipartimento di Fisica e Astronomia ``Augusto Righi''--Universit\`{a} di Bologna, via Piero Gobetti 93/2, I-40129 Bologna, Italy}
\affiliation{INAF - Osservatorio di Astrofisica e Scienza dello Spazio di Bologna, via Piero Gobetti 93/3, I-40129 Bologna, Italy}

\author[0000-0002-7616-7136]{Michele Moresco}
\affiliation{Dipartimento di Fisica e Astronomia ``Augusto Righi''--Universit\`{a} di Bologna, via Piero Gobetti 93/2, I-40129 Bologna, Italy}
\affiliation{INAF - Osservatorio di Astrofisica e Scienza dello Spazio di Bologna, via Piero Gobetti 93/3, I-40129 Bologna, Italy}

\author[0000-0002-4409-5633]{Andrea Cimatti}
\affiliation{Dipartimento di Fisica e Astronomia ``Augusto Righi''--Universit\`{a} di Bologna, via Piero Gobetti 93/2, I-40129 Bologna, Italy}
\affiliation{INAF - Osservatorio Astrofisico di Arcetri, Largo E. Fermi 5, I-50125, Firenze, Italy}

\correspondingauthor{Nicola Borghi}
\email{nicola.borghi6@unibo.it}

\begin{abstract}
	We analyze the stellar ages obtained from a combination of Lick indices in Borghi et al. for 140 massive and passive galaxies selected in the LEGA-C survey at $0.6<z<0.9$. From their median age--redshift relation, we derive a new direct measurement of $H(z)$ without any cosmological model assumption using the cosmic chronometer approach. We thoroughly study the main systematics involved in this analysis: the choice of the Lick indices combination, the binning method, the assumed stellar population model, and the adopted star formation history; these effects are included in the total error budget. We obtain $H(z=0.75)= 98.8\pm33.6~\mathrm{km\,s^{-1}\,Mpc^{-1}}$. In parallel, we also propose a simple framework based on a cosmological model to describe the age--redshift relations in the context of galaxy downsizing. This allows us to derive constraints on the Hubble constant $H_0$ and the typical galaxy formation time. This new $H(z)$ measurement, whose accuracy is currently limited by the scarcity of the sample analyzed, paves the road for the joint study of the stellar populations of individual passive galaxies and the expansion history of the Universe in light of future spectroscopic surveys.
\end{abstract}



\section{Introduction}\label{sec:intro}
	In the era of precision cosmology, independent cosmological probes are fundamental to keep their systematics under control, shed light on the current tensions between different measurements of cosmological parameters, and, ultimately, improve the accuracy of these measurements. Recently, much effort has been devoted to understanding the $\sim4\sigma$ tension between the value of the Hubble constant $H_0$ measured in the local Universe and the one inferred from the cosmic microwave background (CMB) analysis \citep[see][]{Verde2019, DiValentino2021}. If confirmed, this difference would require an extension to the ``vanilla'' 6 parameter $\Lambda$CDM model, introducing new physics at play in the early and/or late epochs. In this context, a cosmological model-independent reconstruction of the expansion history of the Universe can play a crucial role. With the minimal assumption of a Friedmann--Lema\^{i}tre--Robertson--Walker (FLRW) metric, the Hubble parameter $H(z)$ is related to the differential aging of the Universe $\mathrm{d}t_U$ as a function of redshift $z$ by the following equation:
	\begin{equation}\label{eq:cc_Hz}
		H(z) = - \frac{1}{1+z}\frac{\mathrm{d}z}{\mathrm{d}t_U}.
	\end{equation}
	The idea of using a homogeneous population of astrophysical objects to trace $\mathrm{d}t_U$, i.e. \textit{cosmic chronometers} (CC), came from \cite{Jimenez2002}, who proposed massive passively evolving galaxies as ideal CC candidates. Many observational studies revealed that these galaxies build up their mass at high redshift ($z\gtrsim$ 2) over short timescales ($<1$~Gyr) exhausting almost completely their gas reservoir in the very first stages of their life and hence evolve passively to the present age \citep[e.g.,][]{Cimatti2004,Treu2005,Renzini2006,Pozzetti2010,Thomas2010}. However, while redshifts can be measured with high precision (up to 0.1\% with spectroscopic observations), age-dating galaxies is challenged by the complex reconstruction of their star formation history (SFH) and by intrinsic degeneracies within stellar population parameters (e.g., stellar age, formation timescale, and chemical composition, see \citealt{Conroy2013a}).

	In initial works, CCs were selected as red massive galaxies and analyzed with stellar population models to both detect those evolving passively and derive their ages \citep{Jimenez2003, Simon2005, Stern2010}. A different approach was introduced by \citet{Moresco2011}, who proposed to use a direct observable, the spectral break at 4000~\AA\ rest frame (hereafter D4000), to trace the differential age evolution of carefully selected samples of passive galaxies. In fact, the D4000 is linearly correlated (within the considered regimes) with the age of the stellar population, so that $\mathrm{d}z/\mathrm{d}t_U$ can be expressed as $A\times\mathrm{d}z/\mathrm{d\,D4000}$, where the calibration factor $A$ encapsulates stellar population modeling dependencies such as metallicity and star formation history. To date, the majority of the $H(z)$ measurements are based on this method. The other available measurements are made analyzing the ages of luminous red galaxies with the full spectral fitting technique (using, in particular, the \texttt{Ulyss} code; \citealt{Zhang2014,Ratsimbazafy2017}).

	In the previous paper (\citealt{Borghi2022a}, hereafter \citetalias{Borghi2022a}), we derived robust stellar population properties for individual passive galaxies at $z\sim 0.7$, taking advantage of the high-quality spectroscopy of the Large Early Galaxy Astrophysics Census (LEGA-C; \citealt{Wel2016,Straatman2018}) using an optimized set of spectral (Lick) indices. In this Letter, we use the derived age--redshift relation to obtain a new $H(z)$. This enables us, for the first time, to study the stellar population properties (age, metallicity \met, and $\alpha$-enhancement \afe) of a sample of individual cosmic chronometers and use them to constrain the expansion history of the Universe. We also use these data to extract information on $H_0$ and \Om\ jointly with the typical formation time of these systems.

	The $H(z)$ measurement derived in this work does not rely on the assumption of a cosmological model, but only on the minimal assumption of the FLRW metric (Equation~\ref{eq:cc_Hz}). For purely reference values and illustrative purposes, we adopt a `737' cosmology ($H_0=70$~\kmsMpc, $\Om=0.3$, $\OL=0.7$). 

\section{Data}\label{sec:data}
	The present work relies on the detailed stellar population analysis of selected massive and passive galaxies carried out in \citetalias{Borghi2022a}, enabled by the high signal-to-noise ratio $\sim20$ and resolution of $R\sim 3500$ of the LEGA-C DR2 spectra \citep{Wel2016,Straatman2018}. These passive galaxies have a typical stellar velocity dispersion of $\sigma_\star \sim 215$~\kms\ and stellar mass of $\log(M_\star/M_{\odot})\sim 11$. We obtained robust stellar age, metallicity \met, and \afe\ measurements for 140 objects at $0.6\lesssim z \lesssim0.9$, comparing an optimized set of spectral absorption features ($\mathrm{H\delta_A}$, $\mathrm{CN_1}$, $\mathrm{CN_2}$, $\mathrm{Ca4227}$, $\mathrm{G4300}$, $\mathrm{H\gamma_A}$, $\mathrm{H\gamma_F}$, $\mathrm{Fe4383}$, $\mathrm{Fe4531}$, $\mathrm{C_24668}$, hereafter baseline) with the \cite{Thomas2011} models. These models, which we consider for our constraints as oftendone in the literature, assume a single-burst star formation history (SFH). A more realistic SFH is expected to be more complex. However, the analysis of several indicators, including color--color, star formation rate--mass diagrams, and spectroscopic features as the novel Ca \textsc{ii} H/K diagnostic \citep{Moresco2018}, confirm that these galaxies are passively evolving and must have formed over very short time scales (\citetalias{Borghi2022a}). We will quantify the impact of this assumption in Section~\ref{sec:syst} (see also Appendix~\ref{sec:app:SPS}).

	\begin{figure}[t]
		\centering
		\includegraphics[width=\hsize]{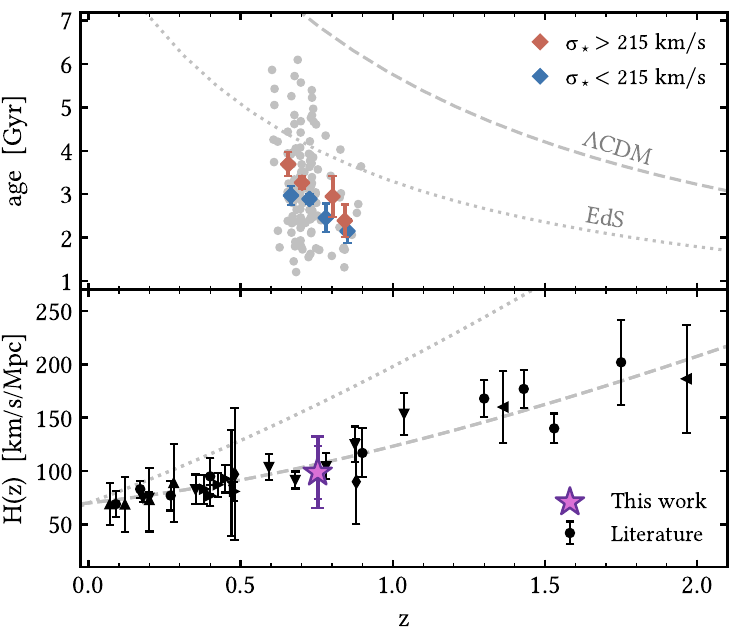}
		\caption{Upper panel: median binned age--redshift relation for 140 passive galaxies analyzed in \citet{Borghi2022a} (gray points) divided into higher (red) and lower (blue) $\sigma_\star$ regimes. Lower panel: $H(z)$ measurement (violet star) with statistical (inner error bar) and total (outer error bar) uncertainties. Black points are literature data from: \citealt{Simon2005} ($\bullet$), \citealt{Stern2010} ($\blacklozenge$), \citealt{Moresco2012b} ($\blacktriangledown $), \citealt{Zhang2014} ($\blacktriangle $), \citealt{Moresco2015} ($\blacktriangleleft $), \citealt{Moresco2016b} ($\blacktriangleright $), and \citealt{Ratsimbazafy2017} ($\blacksquare $). Gray lines are theoretical relations for a standard $\Lambda$CDM (dashed) and Einstein--de Sitter (dotted) cosmology.}\label{fig:Hz_direct} 
	\end{figure}

	\begin{deluxetable*}{lccccccc}[ht!]
		\tablenum{1}
		\tablecaption{Hubble Parameter Measurements. \label{tab:Hz_binned}}
		\tablehead{
		\colhead{Sample} & \colhead{Bins} & \colhead{\# of gal.} & \colhead{$z_\mathrm{eff}$} & \colhead{$\Delta z$} & \colhead{$\Delta \mathrm{age}_{\rm CC}$} & \colhead{$H(z_\mathrm{eff})$} & \colhead{$\sigma_\mathrm{stat}$} \\
		\colhead{} & \colhead{} & \colhead{} & \colhead{} & \colhead{}  & \colhead{(Gyr)} & \colhead{(\kmsMpc)} & \colhead{(\kmsMpc)}  
		}
		\tablewidth{\textwidth}
		\startdata
		Lower $\sigma_\star$  & 1 \& 3 & 20 & 0.723 & 0.114 & -0.514 & 126.3 & 96.4 \\
				  			  & 2 \& 4 & 50 & 0.789 & 0.125 & -0.742 &  92.0  & 36.3  \\
		Higher $\sigma_\star$ & 1 \& 3 & 21 & 0.729 & 0.145 & -0.741 & 111.0 & 80.7  \\
		          			  & 2 \& 4 & 49 & 0.772 & 0.149 & -0.874 & 88.6  & 40.6  \\
		\hline
		Joint     			  & all    & 140  & 0.753\textsuperscript{a} & -- & -- & 98.8\textsuperscript{a}  & 24.8\textsuperscript{a}  \\ 
		\enddata
		\tablenotetext{a}{Joint results are the variance-weighted average of the four Hubble parameter values and are defined at the average effective redshift.}
	\end{deluxetable*}

	\vspace*{-2em}

	The galaxies are divided into two stellar velocity dispersion subsamples using their median $\left\langle \sigma_\star\right\rangle=215~\kms$. For each $\sigma_\star$ regime, we evaluate the median age in four narrow redshift bins (see Figure~\ref{fig:Hz_direct}, upper panel). The constant bin width $\Delta z \simeq 0.075$ corresponds to $\sim 0.4$ Gyr difference in cosmic time, which is also the average age uncertainty. To each bin, we associate an uncertainty computed as the median standard error \citep[NMAD;][]{Hoaglin1983}. The two resulting age--$z$ relations for the higher and lower $\sigma_\star$ regime are approximately parallel and with an offset of $\Delta\age\simeq 0.5$~Gyr. This is consistent with the mass-downsizing scenario, for which more massive galaxies formed earlier and faster. In Section~\ref{sec:app:lowsigma}, we provide further discussion on the inclusion of the lower $\sigma_\star$ population.

	Finally, we stress the utmost importance of avoiding any cosmological prior in the age determination as done in \citetalias{Borghi2022a}; while this kind of prior is commonly used in the literature to reduce the degeneracies between parameters, it is fundamental not to use it in the CC method to avoid introducing circularity in the analysis, with the risk of retrieving the same cosmological parameters adopted as priors.

\section{The direct approach: $H(z)$ measurement}\label{sec:Hz}
	In the cosmic chronometers approach, the Hubble parameter $H(z)$ can be derived directly and without any cosmological assumptions from the differential age evolution of CC, $\Delta\age_{\rm CC}$, within a redshift interval $\Delta z$ (Equation~\ref{eq:cc_Hz}). The quantity $\Delta z/\Delta\age_{\rm CC}$ is measured from the median age--redshift relation between the $i$th and the $i+2$th points for each $\sigma_\star$ subsample, and is defined at an effective redshift of $z_\mathrm{eff}=(z_{i}+z_{i+2})/2$. The choice to use alternate points is to ensure that the evolution in age over the assumed redshift intervals $\Delta z$ ($\sim 0.4$~Gyr of cosmic time) is larger than the statistical scatter, but at the same time sufficiently small to minimize possible systematic effects \citep[see][]{Moresco2012b}. With this bin choice, we obtain four $H(z)$ estimates (Table~\ref{tab:Hz_binned}).

	\begin{figure}
		\centering
		\includegraphics[width=\hsize]{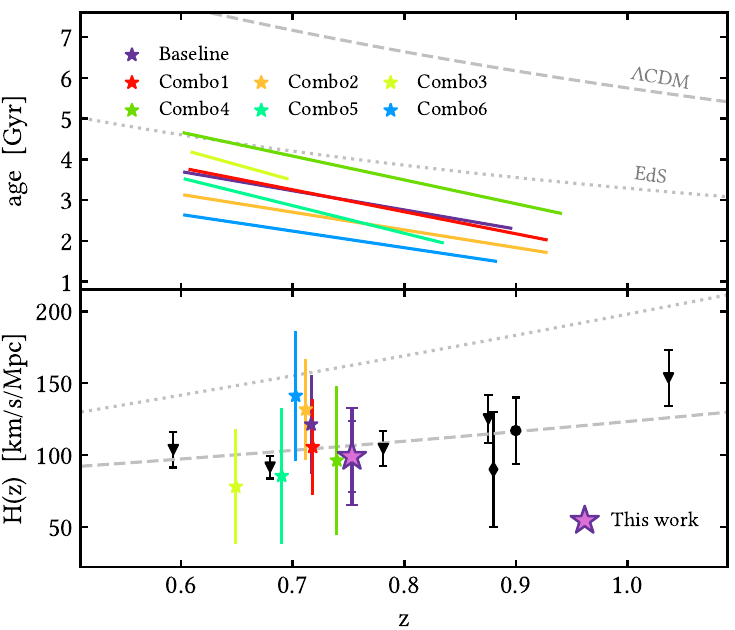}
		\caption{Same as Figure~\ref{fig:Hz_direct}, but showing the results obtained with linear fits to the unbinned age-z relations for different combinations of indices used in the analysis (as presented in \citealt{Borghi2022a}). Gray lines are theoretical relations for a standard $\Lambda$CDM (dashed) and Einstein--de Sitter (dotted) cosmology and are shown for visual inspection purposes only. We note that the EdS model predicts a much flatter slope compared to the ones of the data, which are instead more compatible with a $\Lambda$CDM scenario, as also confirmed the $H(z)$ measurements.} 
		\label{fig:hz_linfit} 
	\end{figure}

	We find that the results for lower and higher $\sigma_\star$ regimes are in very good agreement, with their mean values being within $0.1\sigma$, confirming the idea that these two subpopulations are tracing the same underlying cosmology (see Appendix~\ref{sec:app:lowsigma} for further discussion). Since all four measurements are independent from each other, we combine them using a variance-weighted average, obtaining $H(z=0.75) = 98.8 \pm 24.8~(stat)~\kmsMpc$ at 68\%~C.L. (Figure~\ref{fig:Hz_direct}, lower panel). Our measurement is perfectly consistent with the values estimated with different CC datasets and methods. In particular, the most comparable measurements at this redshift are both from \citet{Moresco2012b} using the D4000 method. Our value lies in between their $H(z=0.68)=91.6\pm 8.0~\kmsMpc$ and $H(z=0.78)=104.5\pm 12.2~\kmsMpc $, differing only by $+0.3\sigma$ and $-0.2\sigma$, respectively. 

	\vspace*{10pt}

	\subsection{Assessing the systematic uncertainties}\label{sec:syst}
		In this section, we explore the main sources of systematic uncertainties in our work. The total uncertainty on $H(z)$ will be computed by adding in quadrature the systematic and statistical contributions.

		\vspace*{-0.5em}
		
        \paragraph{\bf Dependence on the Lick indices set}
			In \citetalias{Borghi2022a}, we studied how different combinations of absorption features impact the derived stellar population parameters. Here, we use this dataset to study the effect of this choice on the age--redshift slope and the final $H(z)$ value. The baseline index combination was devised to maximize the number of indices to be measured given the redshift and wavelength coverage of the various galaxies; moreover, any other index set provides age constraints for fewer objects (down to a dozen for the worst case) and binning them is not always an option. For this reason, to assess this systematic effect, we estimate $\Delta z/\Delta\age_{\rm CC}$ and its associated uncertainty from the inverse slope of the age--redshift relation obtained with a simple linear regression. Results are shown in Figure~\ref{fig:hz_linfit} (see Appendix~C of \citetalias{Borghi2022a} for the indices set definitions).

			Different combinations of indices can provide systematically different absolute age estimates, ranging within $\pm 1$ Gyr. However, we find that the $H(z)$ estimates are consistent with each other and with the more statistically rigorous value obtained with the median binning within $0.4\sigma$, on average. These results clearly highlight the advantages of CC being a differential approach; in other words, the absolute age calibration that might be obtained in different analyses does not significantly affect the final $H(z)$ value, but only the normalization of the age--$z$ relation. 

		\vspace*{-0.5em}

        \paragraph{\bf Dependence on the binning}
			We verify that our result is robust against different redshift binning schemes and adopted estimators. In particular, by using from two up to six redshift intervals, or/and the mean instead of the median age, $H(z)$ results are on average within $0.5\sigma$ with respect to the baseline. We do not use weighted averages because in \citetalias{Borghi2022a} we found that the stellar population analysis intrinsically yields higher uncertainties for older galaxies and this would bias the final age--redshift slope. Finally, we also repeat the analysis using equally populated redshift bins (about 20 objects per bin). This method improves the statistics of single bins at the expense of smaller leverage in redshift. Even in this case, we obtain values in good agreement, with an average difference of $0.5\sigma$.

		\vspace*{-0.5em}

        \paragraph{\bf Dependence on the SPS model}
			The choice of the stellar population model plays a major role in the overall systematics of the CC approach. Quantitatively, \citet{Moresco2020} measured an average contribution of $\sim 7\%$ on the final uncertainty of $H(z)$ using the D4000 method. To assess this effect in our work, we repeat the entire analysis by adopting the $\alpha$-MILES models by \citet{Vazdekis2015}. The detailed analysis is presented in Appendix~\ref{sec:app:SPS:burst}. We find that the $H(z)$ measurements obtained with the assumption of a different SPS model are consistent with the baseline within $0.6\sigma$, on average.

		\vspace*{-0.5em}

        \paragraph{\bf Dependence on a more extended SFH}
			In all the previous analyses, we assume single-burst star formation histories (SFH). This is justified by the accurate selection of passive systems which maximizes the presence of galaxies with coeval SFH concentrated at early cosmic epochs. Here, we assess the effect of assuming a more extended ${\rm SFH}\propto {\rm exp}(t/\tau)$, i.e. exponentially declining with a characteristic timescale $\tau$. The detailed analysis is presented in Appendix~\ref{sec:app:SPS:tau}. As expected from the sample selection, we confirm very short SFHs with a typical $\tau\lesssim 0.4$~Gyr. By analyzing the slopes of the age--redshift relations obtained with these more extended SFHs, we find that the final $H(z)$ measurements differ by only $0.4\sigma$, on average, with respect to the baseline.

			\begin{figure*}[t]
				\centering
				\includegraphics[width=\textwidth]{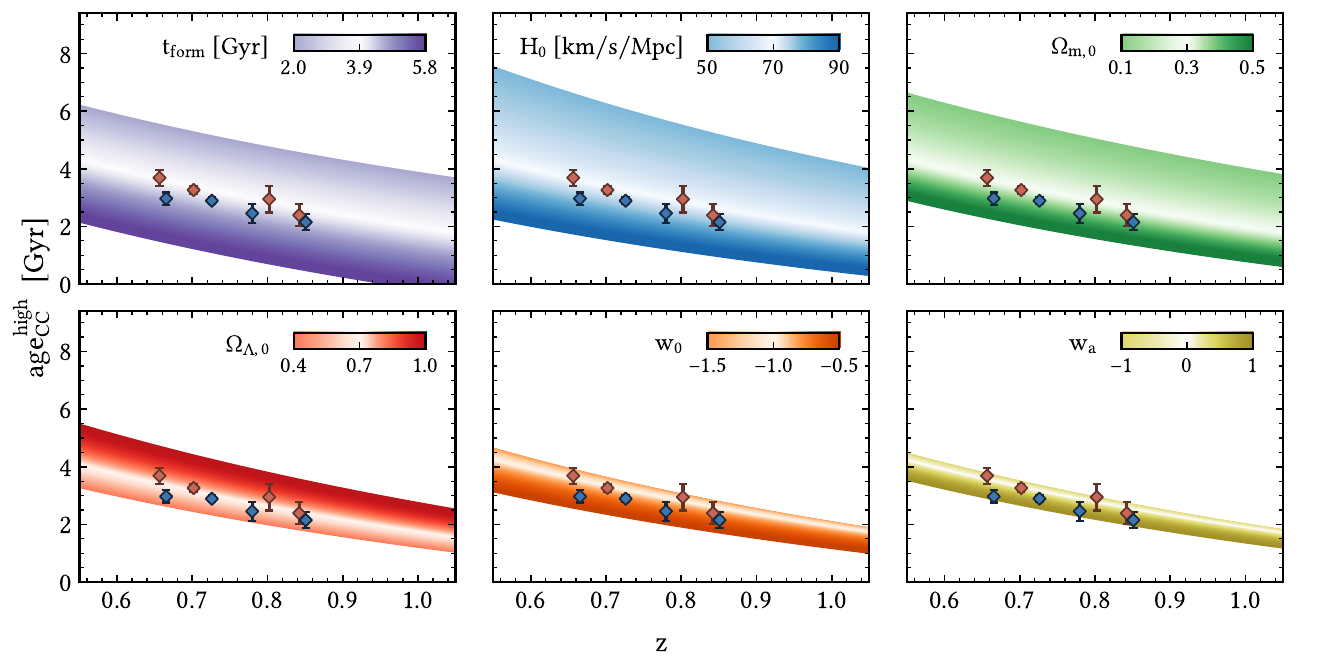}
				\caption{Theoretical age--redshift relations for the high-$\sigma_\star$ subsample of cosmic chronometers. Each panel shows the effect of varying each parameter to the labeled values. Red diamonds are median binned data for the high-$\sigma_\star$ subsample of cosmic chronometers from \citet{Borghi2022a}. For illustrative purposes, we also show the data points of the lower $\sigma_\star$ subsample (blue diamonds), which are about $0.5$~Gyr younger, on average. When not varied, the parameters are set to the following fiducial values: $\tf=3.9$~Gyr, $H_0=70~\kmsMpc$, $\Om=0.3$, $\OL=0.7$, $w_0=-1$, and $w_a=0$.}\label{fig:agez_degeneracies} 
			\end{figure*}

        \paragraph{\bf Final $H(z)$ measurement}
		To summarize, we have collected a total of 15 measurements of $H(z)$ by varying the Lick index set, the redshift binning method, the stellar population synthesis model, and the assumed SFH. With this dataset, we compute a systematic error (obtained from the standard deviation) of $22.7~\kmsMpc$ with respect to our baseline result, where the contribution to the systematic error is almost equally distributed between the various components, being about 1/3 for the Lick index combination, 1/4 for the binning and for the SPS model, and 1/6 for the SFH. This value has been added in quadrature to the statistical error, obtaining as a final result:
		\begin{equation}\label{eq:final_Hz_value}
			H(z=0.75)=98.8\pm33.6\quad\kmsMpc
		\end{equation}
		at 68\% C.L. This is the first $H(z)$ measurement based on the analysis of absorption features of individual passive galaxies, confirming that it is possible to jointly study their stellar population and use the information to derive cosmological constraints. Moreover, the measurement is obtained in a poorly mapped region of the redshift space that is crucial to reconstruct the expansion history of the Universe. In fact, for the assumed fiducial cosmology the transition between a decelerated and accelerated expansion, or transition redshift, occurs at $z_t=0.67$. This promising result, whose accuracy is currently limited by the scarcity of the sample analyzed, has to be seen as a first step toward a detailed study of individual cosmic chronometers in light of future large spectroscopic surveys.

		\begin{figure*}[t!]
			\centering
			\includegraphics[width=0.38\textwidth]{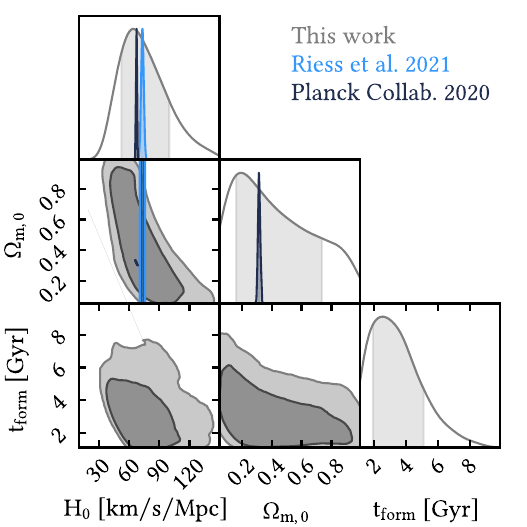}
			\includegraphics[width=0.48\textwidth]{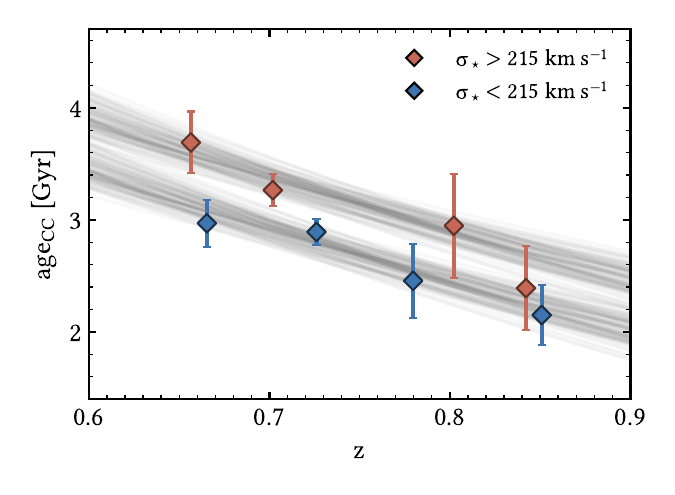}
			\caption{Constraints from age--redshift relations assuming a flat $\Lambda$CDM model. Left: corner plot for $H_0$, \Om, \tf, showing $1\sigma$ (darker shade) and $2\sigma$ (lighter shade) regions. Vertical shaded bands are the cumulative $1\sigma$ confidence regions. Our results are compared with those from \citet{Riess2021} and \citet[][TT,TE,EE+lowE+lensing]{PlankCollaboration2018}. Right: resulting fits (gray lines) to the observed age--redshift relations (diamonds).}
			\label{fig:agez_corner_fLCDM}
		\end{figure*}

\section{Cosmological constraints from the analysis of age--redshift relations}\label{sec:agez}
	The age--redshift relations can be used to set constraints on the Hubble constant $H_0$, and other cosmological parameters \citep[e.g.,][]{Jimenez2019, Vagnozzi2021agez, Krishnan2021}. A recent measurement of $H_0$ in the local Universe has been obtained by \cite{Riess2021} using the luminosity distances of type Ia supernovae calibrated with Cepheid variable stars, $H_0=73.2\pm1.3~\kmsMpc$ at $68\%$ C.L. This value is at 4$\sigma$ tension with the $\Lambda$CDM model-dependent value inferred from the CMB, $H_0=67.36\pm 0.54~\kmsMpc$ \citep[][TT,TE,EE+lowE+lensing]{PlankCollaboration2018}. 

	In this Section, we propose a simple scheme to derive cosmological parameters from the age--redshift relations of different subsamples of CCs binned by their stellar velocity dispersion $\sigma_\star$ in the context of a 
	downsizing evolution (more massive galaxies formed earlier). 

	\subsection{The model}\label{sec:agez:model}
		The age of the Universe as a function of redshift, $t_U(z)$, can be predicted from cosmological models. With the minimal assumption of an FLRW metric:
		\begin{equation}\label{eq:tU_z}
			t_U(z) = \frac{1}{H_0} \int^\infty_z \frac{\mathrm{d}z'}{\left(1+z'\right)E(z')},
		\end{equation}
		where $E(z)$ is the normalized Hubble parameter. Here we assume that the late-time expansion history is described by a flat $w_0 w_a$CDM universe, where the dark-energy equation of state varies with cosmic time under the CPL parameterization, $w(z) = w_0 + w_a(z/(1 + z))$ \citep{Chevallier2001, Linder2003}, therefore:
		\begin{align}
			E(z) &= \sqrt{\Om(1+z)^3+(1-\Om)f(z)}, \\
			f(z) &= (1+z)^{3(1+w_0+w_a)}\mathrm{e}^{-3w_a\frac{z}{1+z}}
		\end{align}
		where radiation is not considered since its contribution is negligible in the late Universe. The function $f(z)$ describes the dark-energy contribution and for a flat $\Lambda$CDM model ($w_0=-1$, $w_a=0$) it becomes $f(z)=1$. 

		Given the inverse relationship between $t_U(z)$ and $H_0$, lower limits on $t_U(z)$ from the ages of the oldest objects would determine upper limits on the local $H_0$ value. Recently, this method has been applied by \citet{Vagnozzi2021agez} to obtain constraints on $H_0$ from galaxies and quasars observed up to $z\sim 8$. 

		Galaxy formation occurs after a time \tf\ from the big bang, which could in principle vary with redshift depending on the considered sample. However, CCs are a population of objects selected to be very coeval in formation time. Therefore, their age--$z$ relation can be written as:
		\begin{equation}\label{eq:Hz_meas}
			\age_\mathrm{CC}(z)=t_U(z)-\tf.
		\end{equation}
		According to the downsizing scenario, galaxy mass is a main driver of galaxy formation and evolution, with more massive galaxies forming their stars at earlier cosmic epochs with respect to less massive ones. For this reason, multiple parallel age--redshift relations for different $\sigma_\star$ populations are expected (and actually visible in the current dataset). Therefore, we use both the lower and higher $\sigma_\star$ subsamples as homogeneous tracers of the age of the Universe by assuming a constant offset in formation time $\Delta t_\mathrm{form}$ computed as the mean age difference. We take as a reference the higher $\sigma_\star$ age--$z$ relation $\age_\mathrm{CC}^\mathrm{high}(z)$, so that $\age_\mathrm{CC}^\mathrm{low}(z)=\age_\mathrm{CC}^\mathrm{high}(z)-\Delta t_\mathrm{form}$. In Figure~\ref{fig:agez_degeneracies}, we illustrate the dependency of $\age_\mathrm{CC}^\mathrm{high}(z)$ on the typical formation time and the cosmological parameters by varying one parameter at a time. 

		As expected from Equation~\ref{eq:tU_z}, similar age--$z$ trends are found by increasing \tf\ (hence $t_U$) and decreasing $H_0$ (and vice versa). A less evident anticorrelation is observed between $H_0-\Om$ and $\Om-\OL$. The latter is orthogonal to the degeneracy that is present in CMB-only data, so that the combination of these two independent probes can eventually provide more stringent constraints on cosmological parameters \citep[see, e.g.,][]{Moresco2016b,Vagnozzi2021eppur}. Finally, it is clear that with the current data it is not possible to set strong constraints on the dark energy equation of state parameters $w_0$ and $w_a$ because of their smaller effect on $\age_\mathrm{CC}(z)$. 

		In our analysis, we therefore assume a flat $\Lambda$CDM universe ($\OL=1-\Om$, $w_0=-1$, $w_a=0$), so that the final model is described by three parameters, $\boldsymbol{\theta}=\left(\tf, H_0, \Om\right)$. We constrain these parameters by using the affine-invariant Markov Chain Monte Carlo sampler \texttt{emcee} \citep{ForemanMackey2019}, assuming a Gaussian likelihood function $\propto e^{-\chi^2/2}$. Priors are set to uniform, noninformative, $H_0~\sim~\mathcal{U}(0, 150)$~\kmsMpc, $\Om~\sim~\mathcal{U}(0.01, 0.99)$, and $\tf~\sim~\mathcal{U}(1, 10)$~Gyr. The final values and associated uncertainties are defined as the cumulative mean and $1\sigma$ values of the marginalized posterior distributions.

	\subsection{Results}\label{sec:agez:results}
		The results are shown in Figure~\ref{fig:agez_corner_fLCDM}. We obtain $H_0=72^{+27}_{-19}$~\kmsMpc, $\Om=0.38^{+0.36}_{-0.23}$, and $\tf=3.2^{+1.8}_{-1.3}$~Gyr. Given the large uncertainties and the small redshift range sampled, our current result is in agreement with both early- and late-Universe $H_0$ determinations; indeed, this method is limited by the intrinsic degeneracies between the parameters shown in Figure~\ref{fig:agez_degeneracies}. We note, however, that these constraints can be significantly improved by increasing the redshift leverage and accuracy of the data, as, for example, could be done by analyzing massive and passive galaxies from proposed spectroscopic missions such as the ATLAS probe \citep{Wang2019}. Differently from the standard CC method presented in Section~\ref{sec:Hz}, the analysis of $\age_\mathrm{CC}(z)$ relies on absolute ages estimates and therefore requires an accurate calibration of galaxies' ages and SFHs and a homogeneous analysis between different samples.

		We also repeat the analysis assuming a Gaussian prior on $\Om\sim\mathcal{N}(0.316, 0.007)$ based on \citep{PlankCollaboration2018} TT,TE,EE+lowE+lensing results. In this case, we obtain $H_0=77^{+20}_{-17}$~\kmsMpc, and $\tf=3.0^{+1.7}_{-1.2}$~Gyr with a significant degeneracy between the two parameters.

\section{Conclusions}\label{sec:conclusions}
	In this Letter, we build upon our previous analysis of stellar population parameters of 140 individual passive galaxies at intermediate redshift \citep{Borghi2022a} to derive cosmological constraints using the cosmic chronometer approach. 
	\begin{enumerate}
		\item We derive a new direct and cosmology-independent estimate of the Hubble parameter $H(z=0.75)=98.8\pm33.6~\kmsMpc$, including both statistical and systematic uncertainties. The latter are obtained by varying the indices adopted to estimate mean stellar ages, the binning scheme, and by assuming different stellar population synthesis models and star formation histories. The accuracy is dominated at the moment by the limited statistics of the sample of cosmic chronometers studied, but nevertheless provide interesting perspectives in light of future large spectroscopic surveys.
		\item We propose a simple model to analyze age--redshift relations of cosmic chronometers at different stellar velocity dispersion $\sigma_\star$ regimes. By assuming a flat $\Lambda$CDM universe, we obtain $H_0=72^{+27}_{-19}~\kmsMpc$ and a typical formation time of $\tf=3.2^{+1.8}_{-1.3}$~Gyr after the big bang for the high $\sigma_\star$ ($>215~\kms$) subsample. In this second approach, it will be crucial to improve the reliability of galaxy's absolute ages using very high-quality spectra combined with up-to-date stellar population models.
	\end{enumerate}
	
	This work demonstrates that it is possible to extend the cosmic chronometer approach by performing a detailed study of the stellar populations of individual galaxies with spectral indices, providing at the same time information on galaxy evolution and cosmology. In view of the extremely interesting constraints to $H_0$ from gravitational waves \citep[e.g., GW170817;][]{Abbott2017} and of the improvements expected in the near future, an important step forward will be the combination of CC and GW analyses to reconstruct for the first time a cosmology-independent measurement of the expansion history of the Universe from $z\sim 0$ to $z\sim2$.

\begin{acknowledgments}
    We thank the anonymous referee for the constructive comments and suggestions that helped to improve this paper. This work is based on data products from observations made with ESO Telescopes at the La Silla Paranal Observatory under program ID 194.AF2005(A-N). We thank the LEGA-C team for making their dataset public. N.B. and M.M. acknowledge support from MIUR, PRIN 2017 (grant 20179ZF5KS). M.M. and A.C. acknowledge the grants ASI n.I/023/12/0 and ASI n.2018-23-HH.0. A.C. acknowledges the support from grant PRIN MIUR 2017 - 20173ML3WW\_001.
\end{acknowledgments}

\software{\textsc{Astropy} \citep{AstropyCollaboration2018},	
		  \textsc{ChainConsumer} \citep{Hinton2016},  
          \textsc{CosmoBolognaLib} \citep{Marulli2016}, 
          \textsc{emcee} \citep{ForemanMackey2019},
          \textsc{Matplotlib} \citep{Hunter2007},
          \textsc{Numpy} \citep{Harris2020},
          \textsc{PyLick} \citep{Borghi2022a}.
          }

\bibliography{references}{}

\begin{appendix}
\vspace*{-4.8em}

\section{Assessing the dependence of the results on the SPS model adopted}\label{sec:app:SPS}
	To verify the dependence of our results on the assumed stellar population synthesis (SPS) model, we repeat the entire analysis by adopting the $\alpha$-MILES models by \citealt{Vazdekis2015} (hereafter \citetalias{Vazdekis2015}). Similarly to \citetalias{Thomas2011}, they are generated with variable \age, \met, \afe\ parameters, and use an updated version of the same empirical stellar library \citep[MILES,][]{FalconBarroso2011}, but are based on corrections from theoretical stellar spectra and assume different stellar isochrones \citep[BaSTI,][]{Pietrinferni2006}. We note that with respect to \citetalias{Thomas2011}, one of the drawbacks of \citetalias{Vazdekis2015} models is that they allow a poorer exploration of the parameter space, having, in particular, a smaller sampling of $\afe=0,0.4$. This introduces some limitations in their use, as will be discussed below, and is one of the reasons why we adopted \citetalias{Thomas2011} models as our reference. However, they give us the possibility to go a step further in the analysis of stellar population properties and test the assumption of a more extended star formation history ($\mathrm{SFH} \equiv \mathrm{SFR}(t)$). In particular, we adopt an exponentially declining function:
	\begin{equation}\label{eq:SFHtau}
		\mathrm{SFR}(t)\propto {\rm e}^{-(\age-t)/\tau},
	\end{equation}
	where $\tau$ is the characteristic star formation timescale. 
	
	The following analysis closely follows the approach adopted in \citetalias{Borghi2022a} to measure indices in the observed data; the reader may refer to section 3 for further details. We generate synthetic spectra\footnote{We use the web tools available at: \url{http://research.iac.es/proyecto/miles/pages/webtools/tune-ssp-models.php} and \url{http://research.iac.es/proyecto/miles/pages/webtools/get-spectra-for-sfhs.php}} with variable \age, \met, \afe, and $\tau$ covering the wavelength range $3550<\lambda/\text{\AA}<5500$ at a resolution of 2.5~\AA\ FWHM and measure the main spectral indices with \texttt{pyLick}.\footnote{Available at: \url{https://gitlab.com/mmoresco/pylick/}} The original grid, spanning the following parameter space: $0.1<\age/{\rm Gyr}<14$ (14 points), $0.01<\tau/{\rm Gyr}<3$ (7 points), $-2.25<\met<0.40$ (7 points), and only two \afe\ points (0 and 0.4), has been interpolated to a resolution of $0.2$ Gyr in \age\ and $\tau$, and $0.02$ dex in \met\ and \afe. This procedure does not introduce significant differences in the resulting parameters. As in the main analysis, we focus on the baseline set of spectral indices ($\mathrm{H\delta_A}$, $\mathrm{CN_1}$, $\mathrm{CN_2}$, $\mathrm{Ca4227}$, $\mathrm{G4300}$, $\mathrm{H\gamma_A}$, $\mathrm{H\gamma_F}$, $\mathrm{Fe4383}$, $\mathrm{Fe4531}$, $\mathrm{C_24668}$, see Section~\ref{sec:data}), which allows us to maximize the number of constrained galaxies. The indices measured on modeled spectra, which are a function of $\boldsymbol{\theta}=(\age, \tau, \met, \afe)$, are compared to the ones measured on the LEGA-C DR2 spectra at 2.5~\AA\ FWHM and corrected to zero velocity dispersion. Specifically, we adopt an MCMC approach using a log-likelihood function $\ln \mathcal{L} = (- 1/2) \sum_i \left(I_i-I^{\mathrm{mod}}_i(\boldsymbol{\theta})\right)^2/\sigma_i^2$ where $I_i$ and $\sigma_i$ refer to the index and its associated uncertainty. In the results presented here, we explore the entire parameter space allowed from the models and - we emphasize here - no cosmological priors are used to derive galaxy ages.

	At the end of this process, we obtain two datasets describing the stellar population properties of the 140 cosmic chronometers using the \citetalias{Vazdekis2015} models:
	\begin{enumerate}
		\item ${\rm V15-SSP}$: single-burst SFH ($\tau\equiv0$);
		\item ${\rm V15-\tau\mbox{-}decl.}$: exponentially declining SFH (Equation~\ref{eq:SFHtau}).
	\end{enumerate}
	In the following section these we will be compared the \citetalias{Thomas2011} results. For the purposes of this study, we are interested in detecting any possible variation in the trends with redshift (which, as described in Equation~\ref{eq:cc_Hz}, is the quantity needed to constrain $H(z)$). In particular, we will discuss percentage differences in ages and absolute differences in \met\ (already expressed in log units) as a function of $z$:
	\begin{align}
	\begin{split}
		\eta &=\age_{\rm V15}/\age_{\rm TMJ11}-1, \\ \Delta &=\met_{\rm V15}-\met_{\rm TMJ11}.
	\end{split}
	\end{align}
	We note here that the interpolation between the two available \afe\ points is not optimal to capture the granularity of this parameter. Indeed, for almost all the galaxies ($>90\%$), we obtain typical $\afe\sim 0$, which is also the grid point nearest to the values obtained with \citetalias{Thomas2011} models ($\sim0.13$~dex). We have also checked that the baseline index combination is not optimal to capture \afe\ variations with the current models. While further analysis with models with denser \afe\ grid points is needed to better understand these differences and study any possible trend of \afe\ with redshift, the analysis of \age\ and \met--which can be strongly degenerate--is sufficient to explore systematic effects on the final $H(z)$ value. 
	
	\begin{figure*}[t]
		\centering
		\includegraphics[width=0.49\hsize]{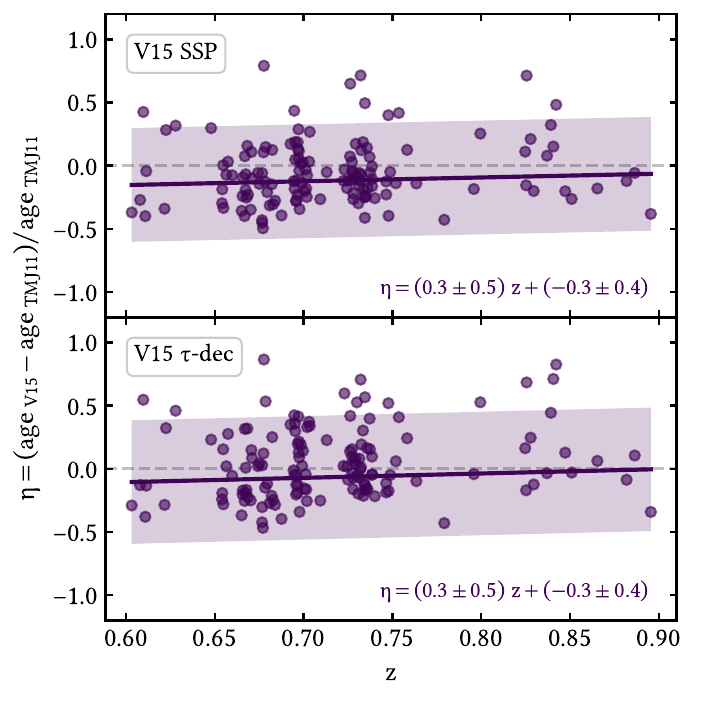}
		\includegraphics[width=0.49\hsize]{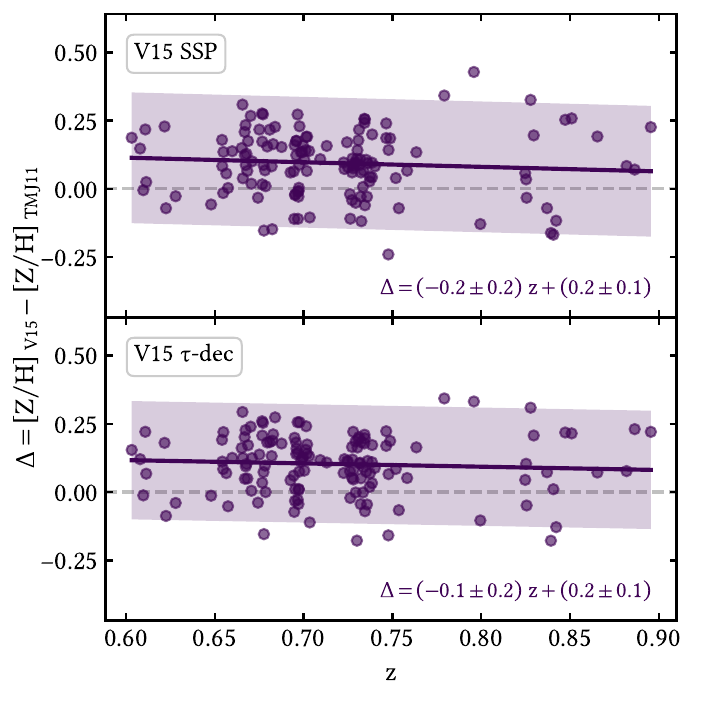}
		\caption{Differences as a function of redshift between stellar ages and metallicities [Z/H] of 140 LEGA-C passive galaxies obtained with \citet{Vazdekis2015} single-burst (SSP, upper panels) and exponentially declining ($\tau$-decl., lower panels) star formation histories versus \citet{Thomas2011} SSP models. Violet lines and shaded regions are robust linear fits and associated $2\sigma$ scatter regions, respectively.} 
		\label{fig:app_differences}
	\end{figure*}

	\subsection{Different model with the same (single-burst) SFH}\label{sec:app:SPS:burst}
		When galaxies are modeled as SSPs, we obtain typical values of $\med{\age}=2.65\pm 0.46$~Gyr and $\med{\met}=0.16\pm 0.27$~dex, differing by $-0.36$~Gyr and $+0.08$~dex, respectively, from the \citetalias{Thomas2011} results. Even if these differences are consistent within $1\sigma$, it is interesting to note that they follow the trend expected from the age-metallicity degeneracy, i.e. younger ages and higher metallicities. However, one of the main advantages of the cosmic chronometer method is that it is insensitive to any systematic offset of the age redshift relation (Equation~\ref{eq:cc_Hz}). We thus explore systematics studying the evolution of differences over redshift (Figure~\ref{fig:app_differences}, upper panels).
 
		It is remarkable that we find no significant deviations as a function of $z$ in the redshift range of interest, with typical differences ranging between $-0.15<\eta<-0.07$ (with 0.22 rms scatter) and $0.07<\Delta<0.12$ (with 0.12 rms scatter). This means that the mean trends of this population of galaxies do not significantly deviate from those observed with \citetalias{Thomas2011} models. As done in the main analysis of this Letter (\S~\ref{sec:Hz}), we compute age--redshift relations for the lower and the higher $\sigma_\star$ subsamples in 4 redshift bins. The final $H(z)$ measurements using median and mean as estimators differ by only $0.6\sigma$ from the baseline. When also testing different binning schemes (including the systematic effects already estimated in Section~\ref{sec:syst}), we obtain measurements consistent within $0.7\sigma$.

	\subsection{Different model with a more extended (exponentially declining) SFH}\label{sec:app:SPS:tau}
		In the main analysis of this Letter, we adopt a single-burst star formation history. Even if our selection criteria were chosen to obtain a sample of galaxies with very short SFH, the single single-burst approximation is not realistic. However, it is important to stress that any constant star formation time-scale for the entire population of these galaxies leads to a vertical shift of the age--redshift relation, therefore, the final $H(z)$ measurement would not be affected. Again, we want to test whether there is any trend of $\tau$ with $z$, which could in principle introduce a bias in the $H(z)$ measurement. At the same time, this analysis gives us the possibility to test how well our assumption of single-burst SFH fits with the observed sample.
		
		\begin{figure*}[t]
			\centering
			\includegraphics[width=\hsize]{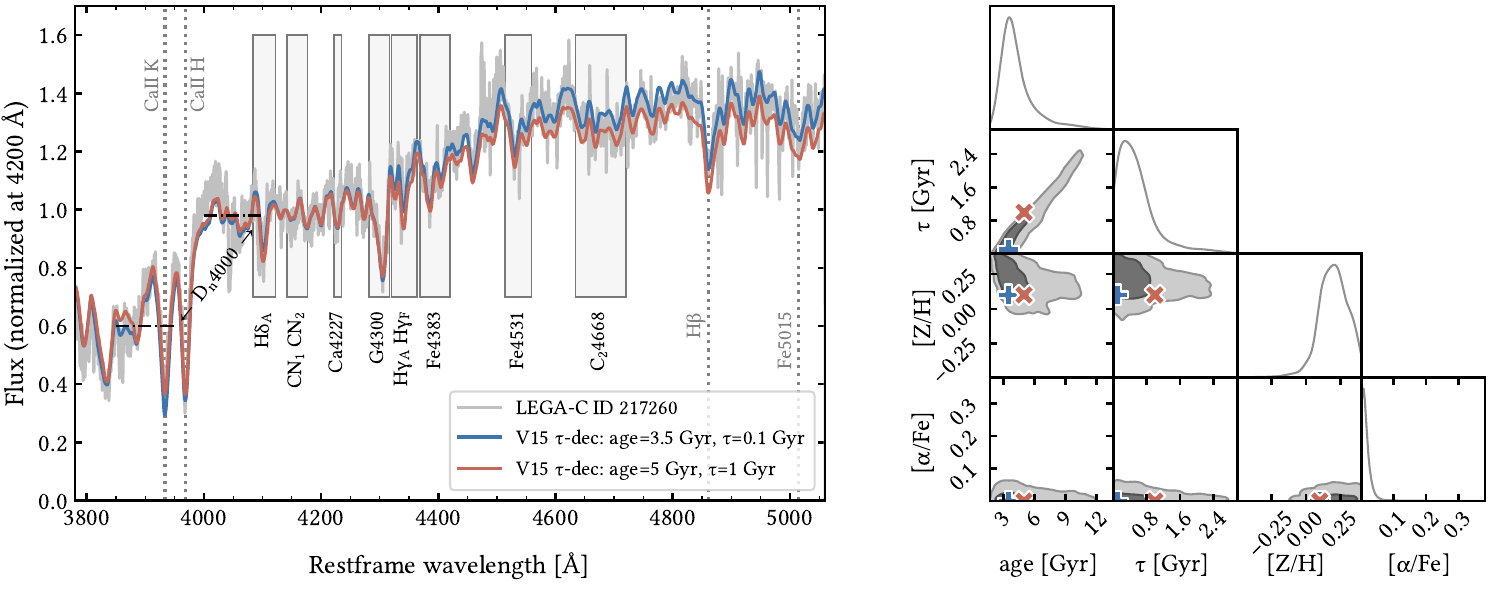}
			\caption{Example of age--star formation timescale degeneracy for a LEGA-C galaxy (ID 217260, ${\rm S/N\simeq22~pix^{-1}}$) comparing the baseline index set (left plot, gray boxes) to the \citet{Vazdekis2015} $\tau$ declining models. Left: observed (gray) and synthetic (blue and red) spectra normalized at 4200~\AA. The blue spectrum corresponds to the model closer to the best-fit parameters, while the red one is taken at the edge of the 1$\sigma$ confidence region. Right: oorner plot for stellar age, star formation timescale $\tau$, stellar metallicity [Z/H], and [$\alpha$/Fe]. The contours enclose $1\sigma$ (darker shade) and $2\sigma$ (lighter shade) confidence regions. Blue and red symbols indicate the values at which the synthetic spectra are generated.} 
			\label{fig:app_degen} 
		\end{figure*}

		Despite the wide range of $\tau$ adopted (0.01--3~Gyr), we obtain typical values of $\med{\tau}=0.24\pm 0.21$~Gyr with only 23\% of the galaxies having $\tau>1$~Gyr. This is an important confirmation that the stellar components of the bulk of these systems formed in very short episodes. We also find no significant dependence on $z$, suggesting minor systematic effects on the final $H(z)$ value. From a more detailed analysis of their posterior distributions, we find that especially systems with $\tau\gtrsim 1$~Gyr suffer from strong degeneracy between \age\ and $\tau$. This degeneracy is well known in the literature \citep[e.g.,][]{Gavazzi2002} and together with the age-metallicity degeneracy is one of the major obstacles in the accurate reconstruction of galaxy star formation histories. Quantitatively, from the analysis of the posterior distributions of our dataset, we find:
		\begin{equation}
			\frac{\Delta \tau}{\Delta \age} \simeq 0.3
		\end{equation}
		i.e., the same set of indices can be reproduced if a galaxy is 1~Gyr older and its star formation time scale extends by 0.3~Gyr. This aspect should be carefully considered when ages from different samples with different SFH assumptions are compared. The age--star formation timescale degeneracy is generally (partially) broken by placing a cosmological prior in the form of an upper limit on galaxies' ages depending on the redshift of observation. However, we shall not use cosmological assumptions in our analysis, as it would introduce a circularity: the retrieved $H(z)$ constraints would be driven by the priors assumed. A possible solution could come from the detailed modeling of Ca \textsc{ii} H and K features, which have proven to be good diagnostics of underlying young stellar populations \citep[see, e.g.,][]{Moresco2018, Borghi2022a}. This can also be seen in Figure~\ref{fig:app_degen} where we show a typical galaxy for which the best fit (blue curve, with $\tau=1$~Gyr and $\age=5$~Gyr) provides similar results with respect to the solution with $\tau=0.01$~Gyr and $\age=3.5$~Gyr (red curve). The spectral indices are insensitive to any difference in the flux normalization, therefore, the normalization of models adopted in the figure (currently chosen in the range $4180<\lambda/\text{\AA}<4220$) is only used for a visual comparison of the models. On the contrary, the difference in the Ca \textsc{ii} H and K lines (not used in this analysis), could be a viable option, preferring the solution with lower $\tau$. This diagnostic will be further explored in future work using the full spectral fitting technique, which allows more flexibility and extensibility to study galaxy SFHs. In this work, we repeat the analysis by fixing an upper prior of $\tau < 0.5$~Gyr corresponding to the upper $1\sigma$ value of the entire population. We verified that this prior does not significantly modify the shape of the age--redshift relation.
	
		We obtain typical values of $\med{\age}=2.88\pm 0.61$~Gyr, $\med{\tau}=0.17\pm 0.09$~Gyr and $\med{\met}=0.21\pm 0.24$~dex, differing by $+0.23$~Gyr, $+0.17$~Gyr, and $+0.05$~dex, respectively, from the SSP results. In Figure~\ref{fig:app_differences} (lower panels) we show the evolution of age and \met\ differences over redshift. 

		Again, it is remarkable that there are no significant deviations as a function of $z$ in the redshift range of interest, with typical differences ranging between $-0.11<\eta<-0.01$ (with 0.24~rms scatter) and $0.08<\Delta<0.12$ (with 0.11~rms scatter). As in the main analysis of this Letter, we have computed median age--redshift relations for the lower and the higher $\sigma_\star$ subsamples in four redshift bins. The final $H(z)$ measurements using median and mean as estimators differ by $0.8\sigma$ from the ones obtained with \citetalias{Vazdekis2015} assuming a single-burst SFH and by $0.4\sigma$ from the baseline (\citetalias{Thomas2011}, single-burst SFH). Also in this case, we test different binning schemes, obtaining measurements consistent within $0.7\sigma$.

	\section{On the inclusion of lower-mass passive galaxies}\label{sec:app:lowsigma}

		According to several studies, the evolution of passive galaxies follows a downsizing pattern, with more massive galaxies forming earlier and faster than less massive ones \citep[e.g.,][]{Renzini2006,Pozzetti2010,Thomas2010}. Therefore, it is important to assess whether the inclusion of the lower $\sigma_\star$ population of passive galaxies in our analysis is well justified and if it could introduce biases in the final $H(z)$ value. In particular, a residual evolution in terms of new stars being formed over cosmic time would result in a flatter age--redshift relation and in a higher $H(z)$ value (Equation~\ref{eq:cc_Hz}). 

		First of all, we stress that the current sample of passive galaxies was carefully selected by combining multiple criteria: photometric NUVrJ cut, spectroscopic emission-line cut (namely, [O \textsc{ii}]$\lambda3727$ and [O \textsc{iii}]$\lambda5007$), and visual inspection of individual spectra to check for residual presence of indicators of ongoing star-formation. Starting from a parent sample of 1622 LEGA-C DR2 galaxies \citep{Straatman2018} we ended up with 140 passive galaxies with age, stellar \met, and \afe\ constraints. In \cite{Borghi2022a}, we split the sample into two bins using the median value ($215~\kms$) as a threshold and found that the two subpopulations (hereafter $S^\mathrm{low}$ and $S^\mathrm{high}$) do not evolve in stellar \met\ and \afe\ within the redshift interval of the study and the stellar populations are consistent with those of their counterparts at $z\approx 0$ under the assumption of a passive evolution. In this section, we further analyze key indicators to test possible biases in tracing the differential age evolution of $S^\mathrm{low}$ within the redshift range $0.6<z<0.9$. We compare them to the results from the analysis of $S^\mathrm{high}$ as a control sample by using the derivative scheme adopted in the baseline analysis (see Section~\ref{sec:Hz}). 

		As a first step, we compute the differential evolution of the 4000\AA\ break (D4000, see Figure~\ref{fig:app_degen}), an age-sensitive index widely adopted in the context of cosmic chronometers \citep[see][]{Moresco2011, Moresco2012b, Moresco2016b}. We find results consistent within $1\sigma$  between the two samples, with ${\rm d\,D_n4000/d}z$ of $-0.5\pm 0.1$ and $-0.6\pm 0.1$ for $S^\mathrm{low}$ and $S^\mathrm{high}$, respectively. A similar conclusion can be drawn from the analysis of the differential evolution of the Ca \textsc{ii} H/K, a diagnostic used to trace recent events of star formation \citep{Borghi2022a}, with ${\rm d\,(H/K)/d}z$ of $0.2\pm 0.1$ and $0.3\pm 0.1$, respectively. To test possible differences in the evolution of stellar \met\ and \afe\ we use the results from the Bayesian analysis of a set of multiple spectral indices (see baseline in Section~\ref{sec:data}). We find a marginal difference in the evolution of \met, with ${\rm d\,\met/d}z$ of $-0.3\pm 0.3$ and $0.1\pm 0.2$ for $S^\mathrm{low}$ and $S^\mathrm{high}$, respectively, and no difference in the evolution of \afe, with ${\rm d\,\afe/d}z$ of $-0.1\pm 0.2$ and $0.0\pm 0.2$, respectively. Finally, we take advantage of the analysis performed in Appendix~\ref{sec:app:SPS} to test possible differences in the differential evolution of the star formation timescale $\tau$, finding ${\rm d\,\tau/d}z$ of $0.1\pm 0.3$ and $-0.1\pm 0.3$, respectively. In conclusion, each test points toward very marginal (if not any) evidence of a slower evolution in redshift for $S^\mathrm{low}$. Therefore, given the limited size of our sample, we decide to include $S^\mathrm{low}$ in the final sample. Future spectroscopic surveys will provide deeper insights on a possible residual evolution in the stellar population of these galaxies.
		
		As a final check, we test whether the inclusion of lower mass passive galaxies biases the final $H(z)$ measurement by separating the contribution of $S^\mathrm{low}$ and $S^\mathrm{high}$. Following the binned analysis presented in Section~\ref{sec:data}, we obtain $H^{\rm lo}(z=0.75) = 101.4\pm34.0~\kmsMpc$ and $H^{\rm hi}(z=0.75) = 96.1\pm36.3~\kmsMpc$. In particular, the value $H^{\rm hi}$ is $5\%$ lower than $H^{\rm lo}$ and only $3\%$ lower than the final $H(z)$ measurement presented in this work (Equation~\ref{eq:final_Hz_value}). We also computed $H^{\rm hi}$ cutting the sample at the $40^{th}$ and $30^{th}$ percentile in $\sigma_\star$, obtaining values always consistent within $0.2\sigma$. 

\end{appendix}

\end{document}